\documentclass[5p,twocolumn,preprint,a4paper,times,12pt]{elsarticle}

\usepackage{amssymb}
\usepackage{wasysym}   
\usepackage{hyperref}

\journal{J. Phys.: Conf. Ser.}

\begin{document}

\begin{frontmatter}

\title{Underground neutron events at Tien~Shan and the properties of~the~$10^{14}-10^{17}$\,eV~EAS muonic component}

\author[a]{A.\,Shepetov}
\cortext[mycorrespondingauthor2]{Corresponding author}
\ead{ashep@www.tien-shan.org}

\author[a]{A.\,Chubenko}

\author[a,b]{O.\,Kryakunova}

\author[e]{O.\,Kalikulov}

\author[a]{S.\,Mamina}

\author[e]{K.\,Mukashev}

\author[a]{R.\,Nam}

\author[a]{V.\,Piscal}

\author[a]{V.\,Ryabov}

\author[d]{T.\,Sadykov}

\author[e]{N.\,Saduev}

\author[a,b]{N.\,Salikhov}

\author[d]{E.\,Tautaev}

\author[a]{L.\,Vildanova}

\author[b]{Zh.\,Zhantayev}

\author[a]{V.\,Zhukov}

\address[a]{P.\,N.\,Lebedev Physical Institute of the Russian Academy of Sciences (LPI), Leninsky~pr., 53, Moscow, Russia}

\address[b]{Institute of Ionosphere, Kamenskoye~plato, Almaty, Kazakhstan}

\address[e]{Al-Farabi Kazakh National University, Institute for Experimental and Theoretical Physics, al-Farabi~pr., 71, Almaty, Kazakhstan}

\address[d]{Satbayev University, Institute of Physics and Technology, 050032, Ibragimova str. 11, Almaty, Kazakhstan}

\begin{abstract}
The events of multiple neutron production under 2000\,g/cm$^2$ thick rock absorber were studied at the Tien~Shan mountain cosmic ray station, at the altitude of 3340\,m above the sea level. From comparison of the experimental and Geant4 simulated neutron multiplicity spectra it follows that the great bulk of these events can be explained by interaction of cosmic ray muons with internal material of the neutron detector. In synchronous operation of the underground neutron monitor with the Tien~Shan shower detector system it was found that the characteristics of the muonic component of extensive air showers which is seemingly responsible for generation of the neutron events underground do change noticeably within the energy range of the knee of primary cosmic ray spectrum. Some peculiar shower events were detected when the neutron signal reveals itself only $\sim$(100--1000)\,$\mu$s after the passage of the shower particles front which probably means an existence of corresponding delay of the muon flux in such events.
\end{abstract}


\end{frontmatter}

\section{Introduction}

Investigation of the events of multiple neutron production in underground detectors of the Tien~Shan mountain cosmic ray station was started about a decade ago. Phenomenological results which have been obtained then on the properties of these events were reported in \cite{undgour1, undgour2,undgour2008icrcmexico}; in these publications it was stated that the original nature of neutron events observed underground still remains unclear. After completion of modification period of the complex detector system for the cosmic ray studies at Tien~Shan station \cite{ontien-nim2016}, and after systematic introduction of the Geant4 package based simulation methods for determination of the properties of its particle detectors it became possible to return to the problem of underground neutron events at new stage of experimental technique. In particular, the detection of neutron events in strict synchronization with the shower installation now permits to study precisely the neutron bearing properties and temporal characteristics of the penetrative component of extensive air showers (EAS). An overview of the results newly obtained in simultaneous operation of the Tien~Shan shower detector system with the  underground neutron detector is subject of the present message.

\section{The underground neutron monitor}

\begin{figure*}
\includegraphics[width=0.6\textwidth, trim=0mm 0mm 0mm 40mm]{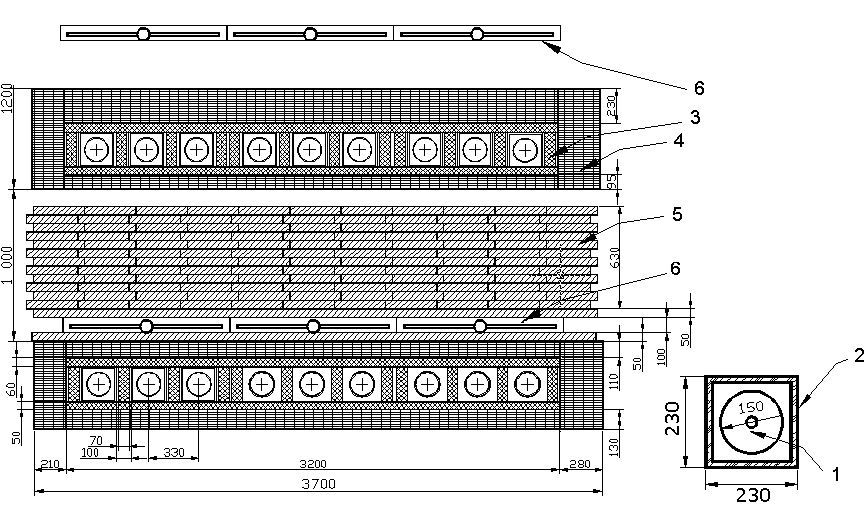}
\includegraphics[width=0.4\textwidth, trim=0mm 0mm 0mm 0mm]{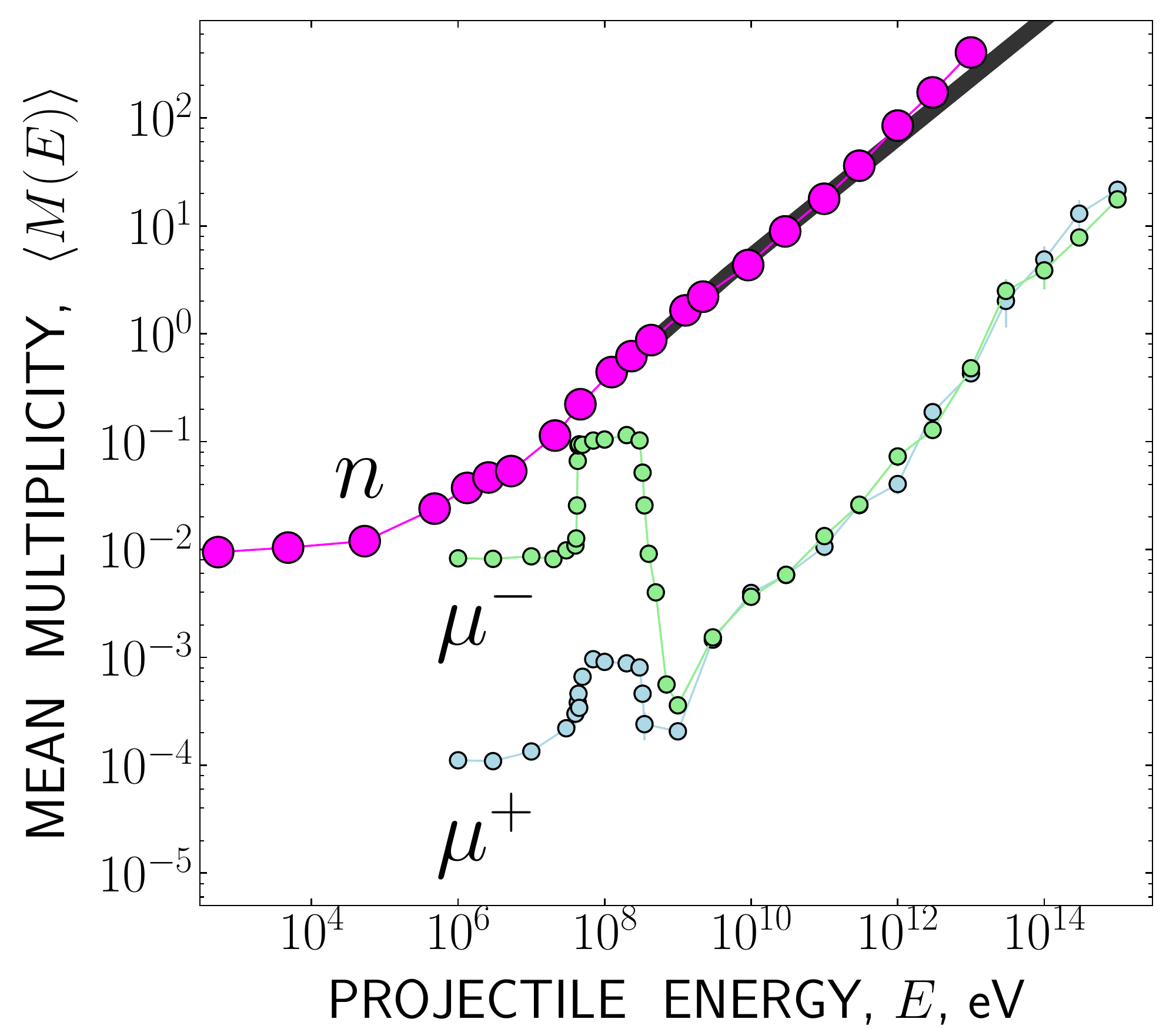}
\caption{\label{figiundgeffici}Left: internal arrangement of the underground neutron monitor installation in the present time: \textit{1}~--~neutron counter, \textit{2}~--~moderator, \textit{3}~--~lead target, \textit{4}~--~rubber (outer~moderator and shielding), \textit{5}~--~iron absorber, \textit{6}~--~plastic~scintillators. Dimensions are shown in millimeters. Right: Geant4 simulation results for the multiplicity of the neutron counters signals of a single monitor unit in dependence on the energy of incident neutrons and muons; continuous straight line beside the neutron data indicates an approximation of the corresponding experimental results from the work \cite{nmn2003}.
}
\end{figure*}

The data discussed further on were obtained in the years 2012--2018 in a long term experiment with the neutron detector which is placed beneath a 2000\,g/cm$^2$ thick absorber, at altitude of 3340\,m above the sea level in the underground room of the Tien-Shan mountain cosmic ray station. Presently, this detector consists of a pair of separated units---\textit{UPPER} and \textit{LOWER}, both of which were made resembling the standard NM64 type neutron supermonitor \cite{carmichel_supermonitor}. In the underground room these detector units, which thus can be referenced as the the \textit{underground monitor}, are placed one above the other as it is shown in the left picture frame of figure~\ref{figiundgeffici}. Both units include the layers of heavy target absorber where penetrative particles of cosmic radiation can experience nuclear interaction with lead nuclei. Evaporation neutrons which originate as a result of this interaction can be detected by the big $\diameter 150\times 2000$~mm$^2$ neutron counters with enriched $^{10}$BF$_3$ gas filling, so the detection of low-energy neutrons is possible there due to the reaction $n(^{10}$B,$^7$Li$)\alpha$. Before detection, the neutrons loose their initial MeV-order kinetic energy down to thermal level in multiple interactions with light nuclei within the sheets of internal moderator material which consists of the wooden boxes surrounding the counters. Another sheets of hydrogen enriched rubber (C$_2$H$_2$) which cover all the unit from outside play the role of external shielding to prevent the influence of environmental low energy neutrons background on the measurement of the cosmic ray connected neutron signal.

Hence, the response of the neutron monitor unit to interaction of a nuclear-active cosmic ray particle is connected with a number of electric pulse signals obtained from its counters during some fixed \textit{time gate} period after this interaction. Hereafter, sum number of such signals will be designated as the neutron \textit{multiplicity} $M$. In \cite{nmn2003} it was shown experimentally that for a monitor unit of considered construction the multiplicity $M$ is nearly proportional to the square root of energy deposit in primary interaction (in the GeV order energy range), and the typical duration of the gate time can be of the order of a few milliseconds. More precisely, the energy dependence $M(E)$ can be defined through the detailed simulation of primary interaction and subsequent moderation and diffusion processes of originating neutrons within the monitor material which can be made on the basis of the modern Geant4 toolkit \cite{yanke2011}. The results of such calculations for the hadronic (neutron) and muonic type primaries are presented in the right plot of figure~\ref{figiundgeffici}.

\begin{figure*}
\begin{minipage}{0.465\textwidth}
\includegraphics[width=\textwidth, trim=5mm 0mm 0mm 0mm]{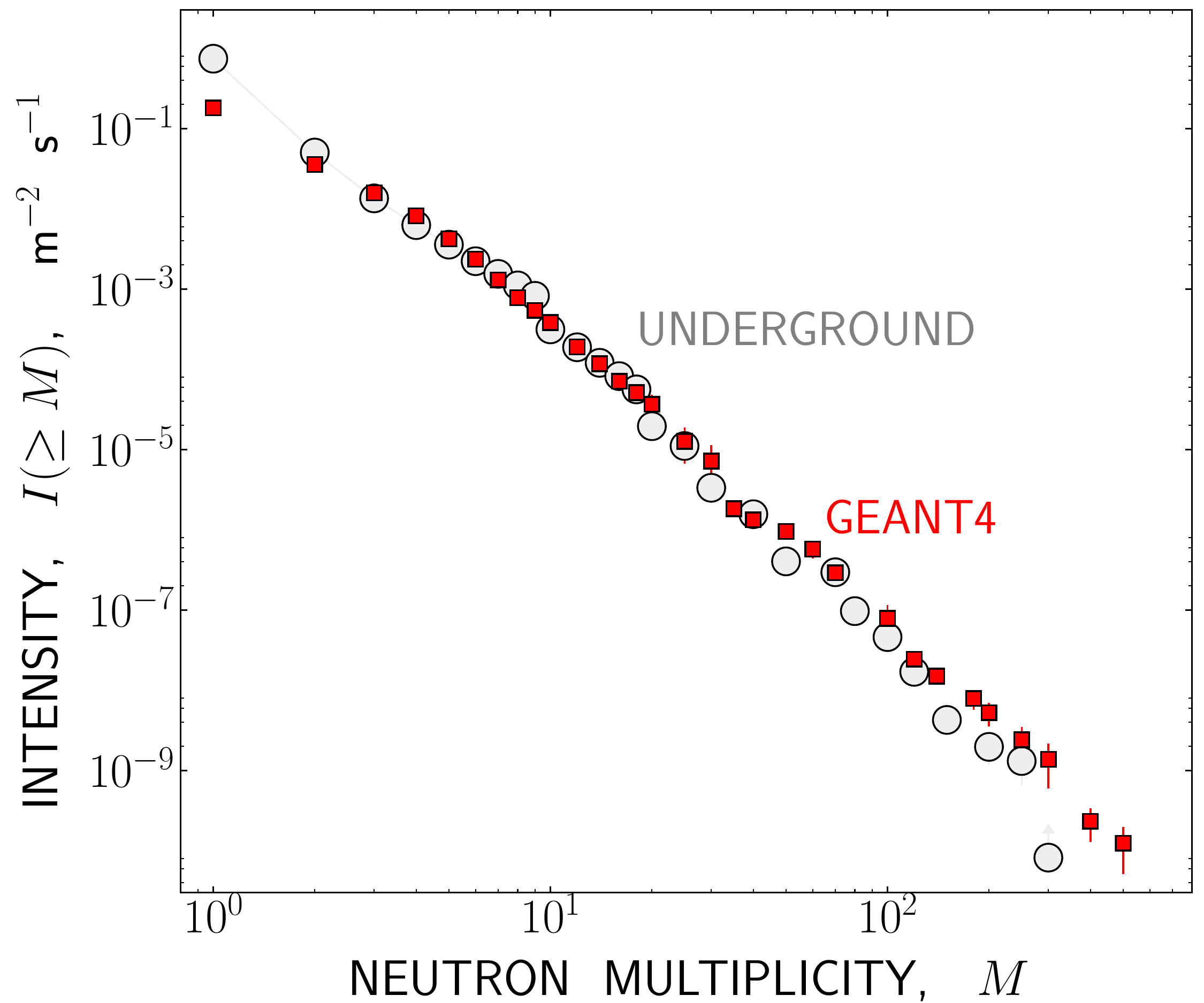}
\caption{\label{figiundgsimispe}Experimental (circles) and Geant4 simulated (squares) multiplicity spectra of neutron events in the underground monitor.}
\end{minipage}
\hspace{2pc}
\begin{minipage}{0.465\textwidth}
\includegraphics[width=\textwidth, trim=5mm 0mm 0mm 0mm]{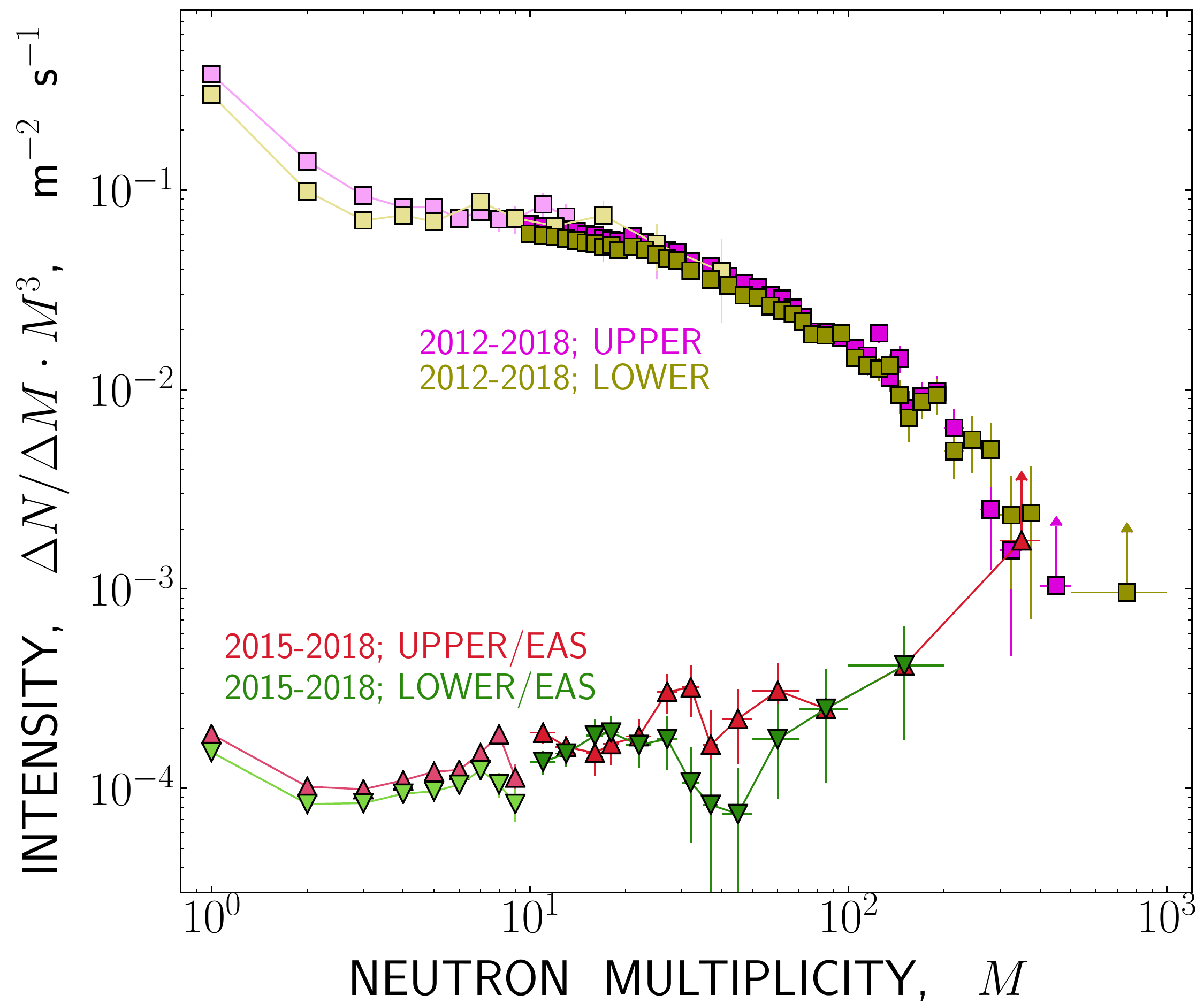}
\caption{\label{figiundgeasspe}Multiplicity spectra of all neutron events seen in the \textit{UPPER} and \textit{LOWER} units of the underground neutron detector (above), and of the events with EAS accompaniment only (below).}\end{minipage}
\end{figure*}

Same Geant4 simulation model can be of use to answer the question which was put in \cite{undgour1} on nature of the underground neutron events. Taking into consideration the spectrum of energy deposits from penetrative cosmic ray particles which has been measured earlier in the same underground room of Tien~Shan station in experiments with the ionization calorimeter \cite{ontienmuons_erlykin1973}, and using it as the input for the Geant4 simulation series with $\mu^\pm$~type primaries one can obtain the expected neutron multiplicity spectrum of underground events. A comparison of such simulated and experimentally measured multiplicity spectra is made in the plot of figure~\ref{figiundgsimispe} where it is seen that both spectra do agree rather well with each other. From this fact a conclusion can be drawn that in the case of underground monitor we deal mostly with the products of nuclear interaction caused by cosmic ray muons, so this monitor as a whole can be used as a specific detector of the cosmic ray muonic component, and particularly of the muons which accompany the passage of extensive air showers.

\section{Underground neutron events and extensive air showers}

\begin{figure*}
\begin{center}
\includegraphics[width=0.49\textwidth, trim=5mm 0mm 20mm 0mm]{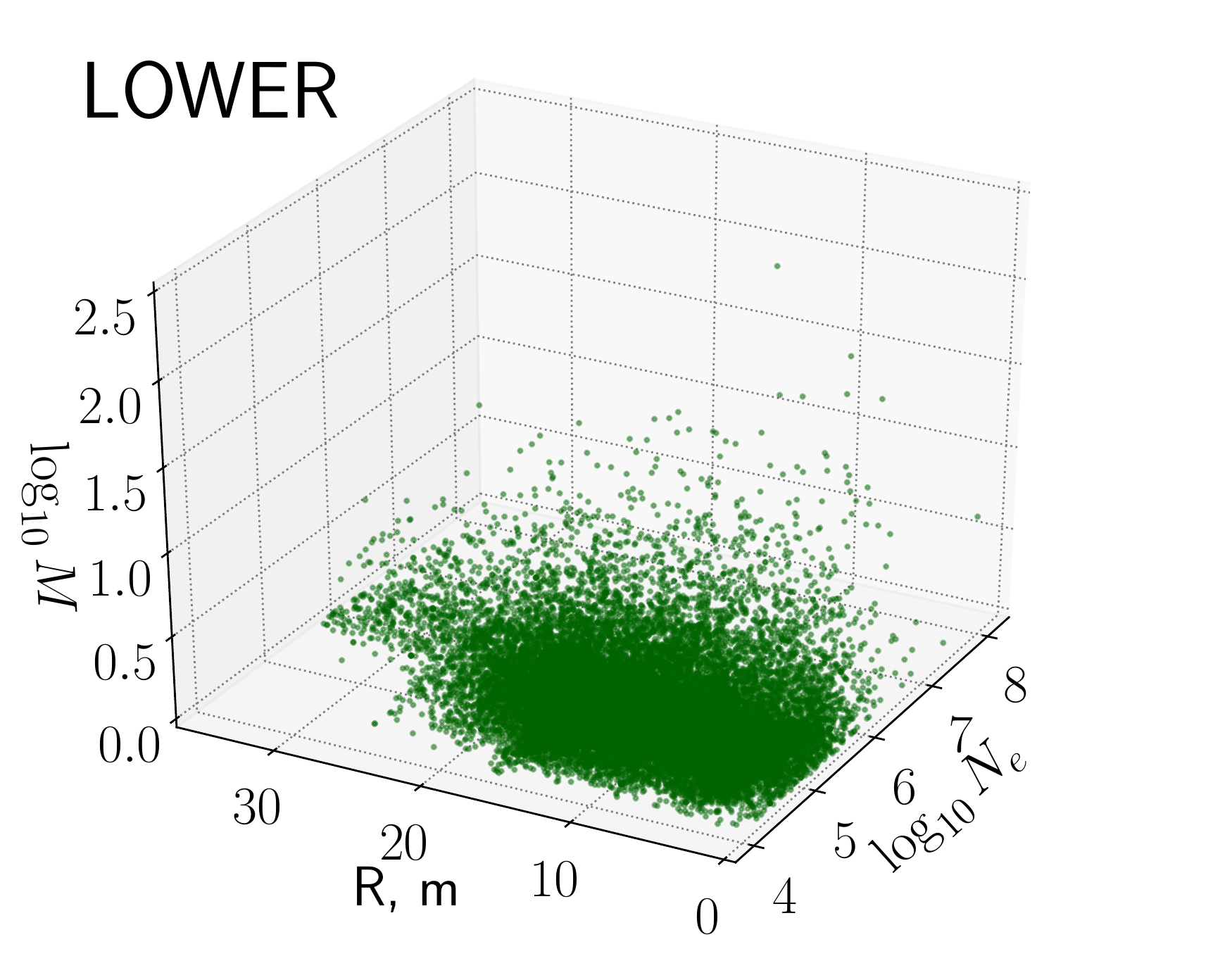}
\includegraphics[width=0.49\textwidth, trim=0mm 0mm 29mm 0mm]{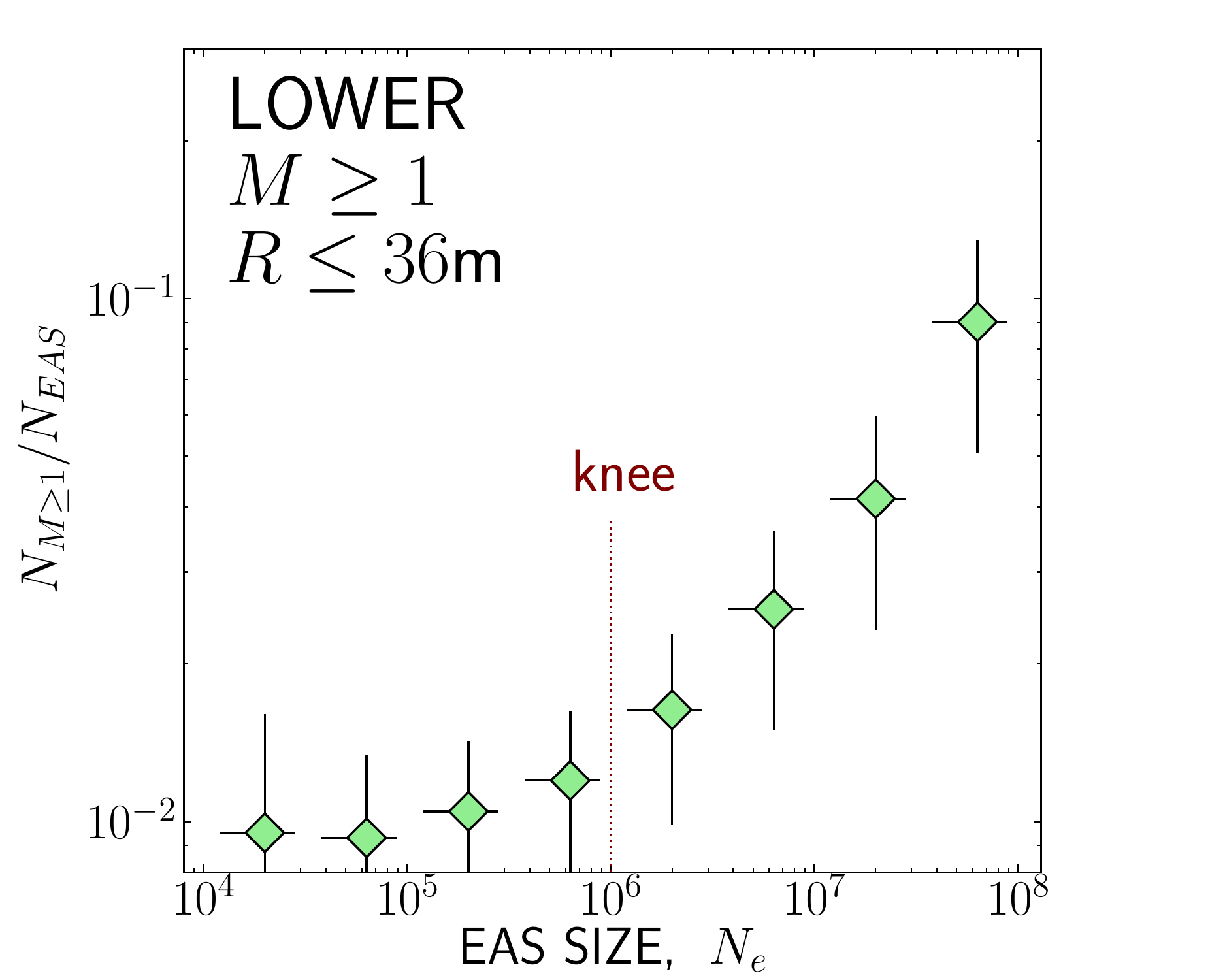}
\caption{Left: correlation plots between the multiplicity of the underground neutron events $M$, the shower size $N_e$, and the core distance $R$ of accompanying EAS. Right: relative share of the EAS with detected neutron signal from muonic interactions underground.
(Here, there are plotted only the data which have been obtained for the lower unit of the underground monitor; those for its upper unit are quite analogous to presented ones).
}
\label{figiundgeascounts}
\end{center}
\end{figure*}

During operation periods of the Tien~Shan shower installation \cite{ontien-nim2016} in the years 2015--2018 the data taking process at the underground neutron monitor was fulfilled under the control of external trigger signal which was generated in the moments when an extensive air shower had been detected above the surface of Tien~Shan station. The trigger generation algorithm  ensured a nearly (90--95)\% selection probability of the EAS with primary energy above $E\gtrsim 10^{15}$\,eV whose axes were coming  through the central part of shower installation, at a distance of $R\leqslant(25-30)$\,m from the underground detector system. The detection of smaller EAS up to $E_0\approx 10^{14}$\,eV was possible as well at that time with somewhat reduced efficiency. The upper energy limit of detected EASs is defined mostly by the geometrical area of the shower detector installation, and by the total duration time of the measurements. For the dataset presented here its value is about $10^{17}$\,eV.

The multiplicity spectra of the neutron events of muonic origin which have been registered simultaneously with an EAS are presented in figure~\ref{figiundgeasspe}, in comparison with analogous spectra calculated over the whole set of underground events. It is seen there that generally the intensity of EAS accompanied events occurs being 2--3~orders of magnitude below the total flux of the $M\lesssim100$ events, but tends to match with the latter in the range of extremely high multiplicities. Seemingly, this difference can be explained by particulars of the shower trigger elaboration threshold and their influence on registration probability of the EAS connected neutron events underground.

To clarify more precisely what are the favorable conditions for detection of the muonic EAS component by neutron signal from its nuclear interaction, a correlation plot can be built between the multiplicity of underground neutron events $M$, and the core distance $R$ and the size (total number of the charged particles) $N_e$ of accompanying showers. An example of such correlation is presented in the left plot of figure~\ref{figiundgeascounts}. The range of distance values here is limited from above with geometrical sizes of the central detector ``carpet'' of the Tien~Shan shower installation ($R^{max}\simeq 36$~m, see \cite{ontien-nim2016}) which was used for precise location of the EAS axis in the measurement series when the considered experimental data were obtained. From correlation plots of figure~\ref{figiundgeascounts} it is seen that the events with non-zero multiplicity values were observed mostly in the cases when a shower axis had passed in relative vicinity ($R\lesssim 20-25$~m) to the projection of the neutron detector position to the surface of the ground, but the distance between the EAS core and the location of the monitor in such events can be somewhat bigger for the showers with $N_e\gtrsim 10^6$. As well, the events with the comparatively high values of detected neutron multiplicity ($M\gtrsim10-30$) were met only amongst the EAS with $N_e\gtrsim 10^6$.

The dependence between the relative amount of EAS events with non-zero multiplicity of detected underground neutron signals and the average size of accompanying shower $N_e$ is illustrated by the right plot of figure~\ref{figiundgeascounts}. In this plot the average number of shower events with $M\geqslant 1$ is normalized to the total statistics of registered EAS with given $N_e$, $\mathcal{N}_{EAS}$. As it follows from this picture, in the showers with $N_e\lesssim 10^6$ the share of $M\geqslant 1$ events remains at an approximately one and the same low level. The latter nearly fits zero, and virtually it can be explained by random coincidences of shower trigger with background neutron events in the monitor. Contrary to this, in the range of shower sizes $N_e\approx 10^6$ the relation $\mathcal{N}_{M\geqslant 1} / \mathcal{N}_{EAS}$ starts to grow, and evidently it can not be put here into any agreement with its previous behaviour.

The revealed sharp increase in relative share of the showers which were accompanied by the underground neutron events means corresponding rise in the average multiplicity, or in the energy, or in both of the EAS connected muons which constitute the original source for generation of these events. As it follows from figure~\ref{figiundgeascounts}, such a change of the average shower characteristics resides somewhere in vicinity of the $N_e\approx 10^6$ EAS size value.

It should be noted that at altitude of the Tien~Shan mountain station the range of EAS sizes $N_e\approx 10^6$ corresponds to the primary energy of a cosmic ray particle $E_0\approx 3\cdot 10^{15}$~eV \cite{ontien_icrc1987__e0_through_ne_ru}, i.\,e. to position of the well-known knee in the energy spectrum of cosmic rays. Since the newly found deviation of the mean characteristics of EAS connected muonic component occurs just at the same point on energy scale, probably it could be added to a wide list of the various peculiar effects which have been met up to date within this energy range by many research groups, and particularly in the experiments which have been held previously at the Tien~Shan cosmic ray station \cite{ontien-nim2016}.

\begin{figure*}
\begin{center}
\includegraphics[width=0.49\textwidth, trim=0mm 10mm 20mm 0mm]{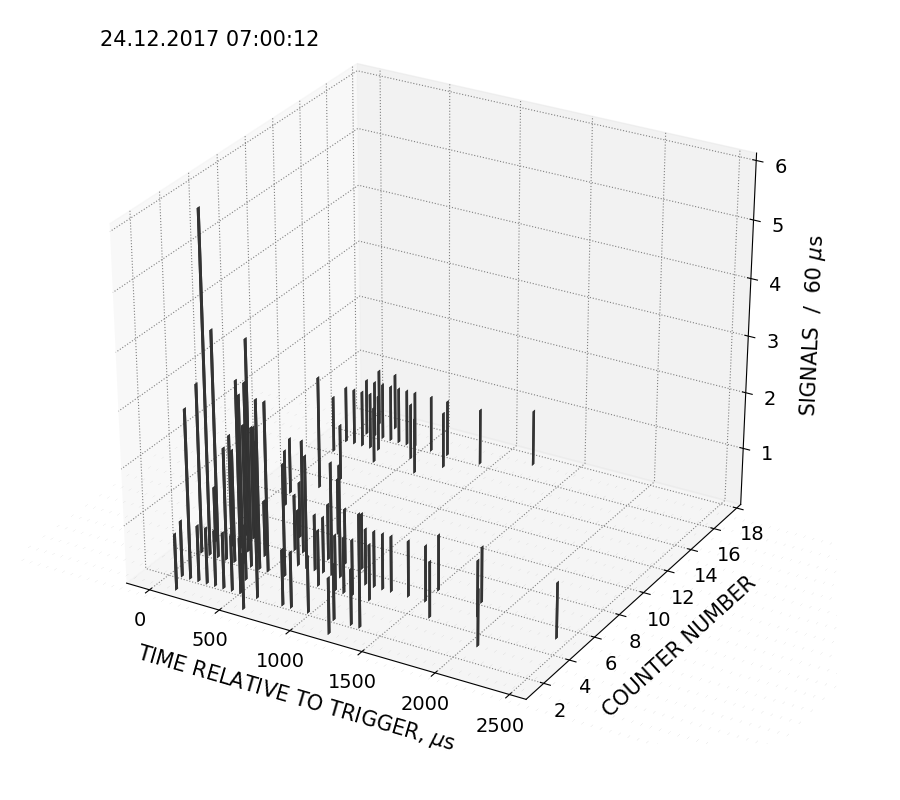}
\includegraphics[width=0.49\textwidth, trim=0mm 10mm 20mm 0mm]{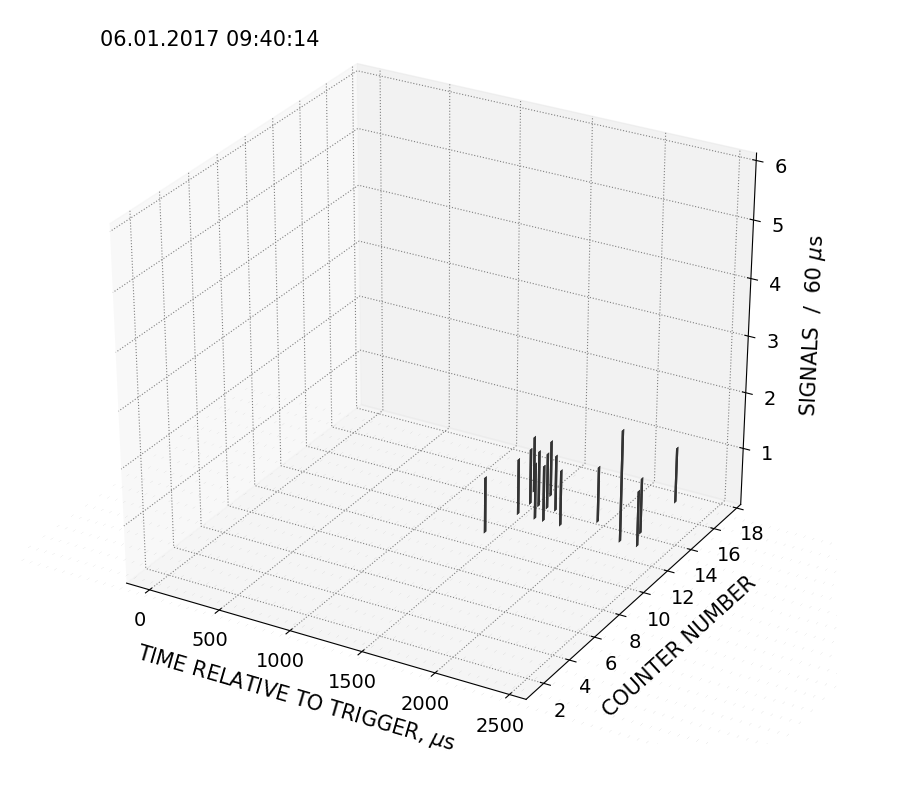}
\\
\includegraphics[width=0.49\textwidth, trim=0mm 20mm 0mm 0mm]{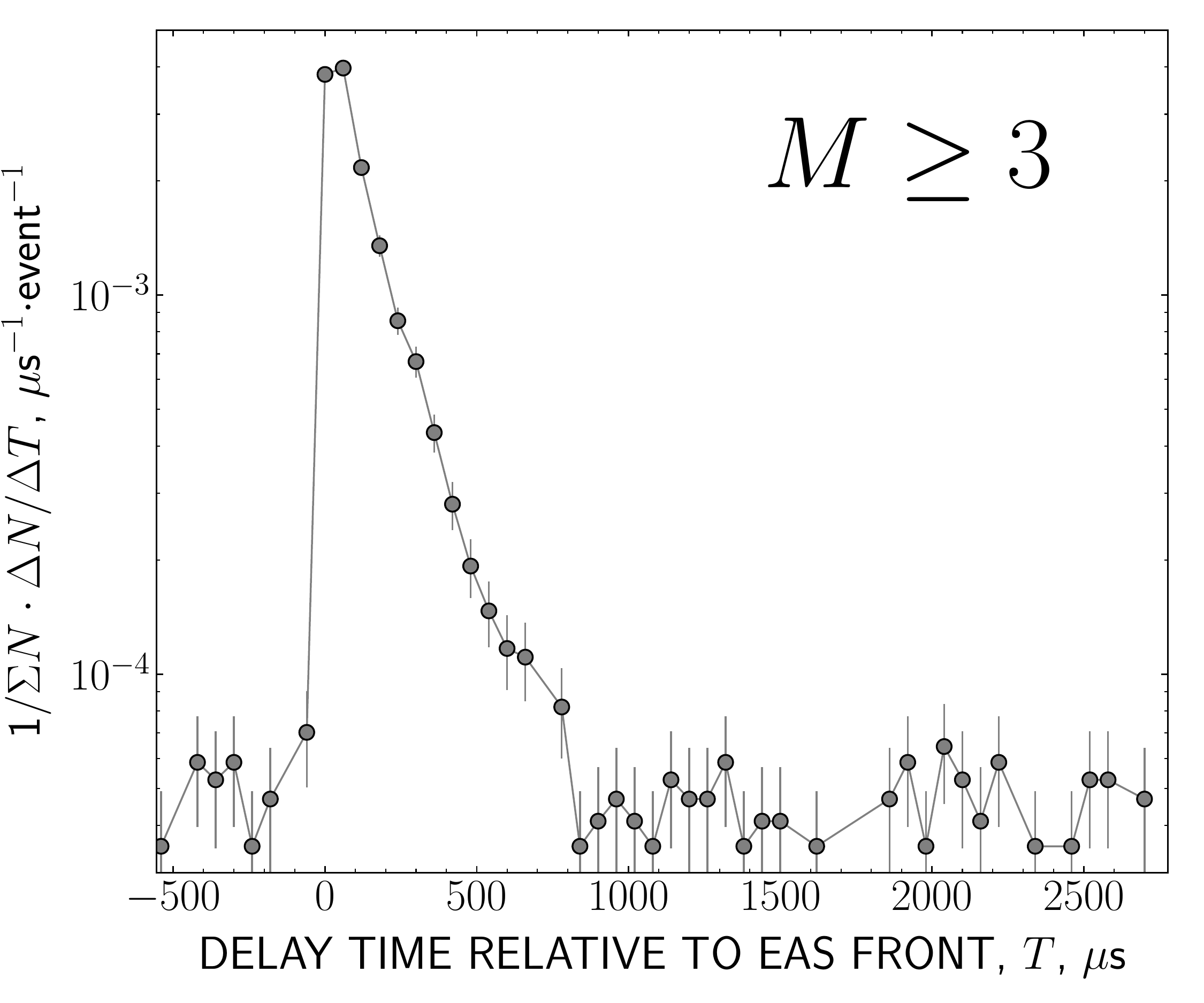}
\includegraphics[width=0.49\textwidth, trim=0mm 20mm 0mm 0mm]{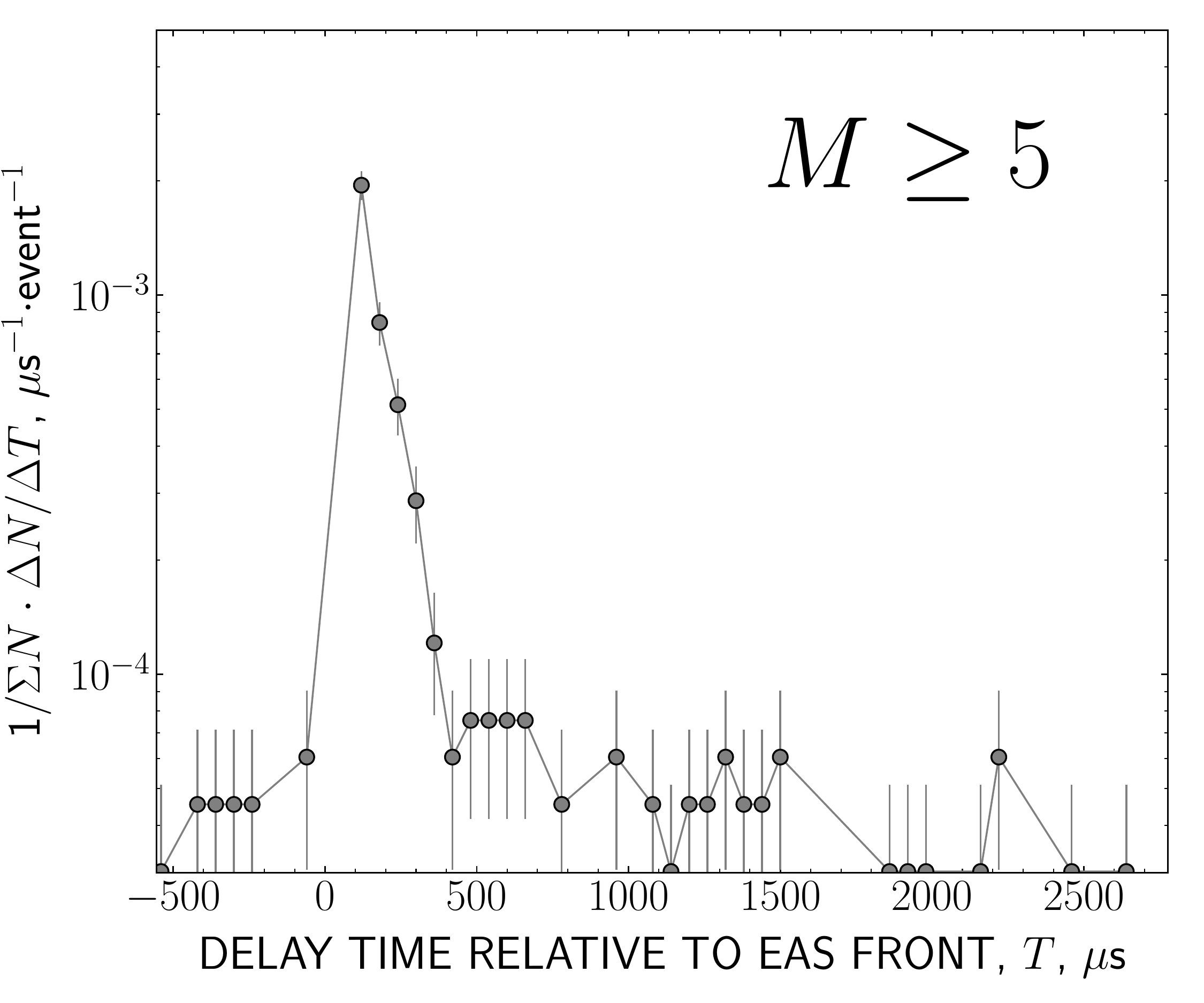}
\end{center}
\caption{Top: a sample of normal (left) and delayed (right) time distributions of neutron signal in EAS connected events of underground monitor.  Zero point of  time axis in both plots corresponds to the moment of EAS trigger; vertical axis is graduated in the units of neutron counter pulses which have come during a single 60~$\mu$s long time interval. Neutron counters 1\,\ldots9 correspond to the upper monitor unit, 10\,\ldots18---to the lower one. Bottom: distribution of the time delay values between the shower front and the beginning of the neutron signal development in underground events with sum multiplicity $M\geqslant 3$ and $M\geqslant 5$.}
\label{figiundgeasdelays}
\end{figure*}

Another peculiarity observed in behaviour of the underground neutron events concerns their temporal characteristics. Since the shower trigger provides a rather strict binding to the EAS passage it is possible to trace precisely the time distribution of subsequent neutron signals with regard to initial moment of nuclear interactions which take place within detector unit at the hitting moment of EAS muons. It is natural to expect that generally such distributions must be of an exponent type which results from diffusion of the newly born evaporation neutrons within the monitor material. Indeed, most part of detected events do have such exponential shape, like a sample event shown in the top left panel of figure~\ref{figiundgeasdelays}. At the same time, among all EAS connected events it was found a noticeable amount of cases when the beginning of exponential intensity decrease has been too late by a some considerable time $T$ in relation to the moment of shower trigger, so that $T$ can be of up to a few hundreds of microseconds order. An example of a such ``delayed'' event is presented in the top right panel of figure~\ref{figiundgeasdelays}.

Of course, certain part of the uncommon delays seen underground can be explained by arbitrary overlapping between the moments of EAS trigger and the background neutron events in monitor. To elucidate the role of  random coincidences in the discussed effect a distribution can be built over observed values of the delay times $T$. Such distribution (normalized to the total amount of registered EAS cases with a non-zero neutron accompaniment underground) is presented in two bottom plots of figure~\ref{figiundgeasdelays} which were drawn for the events with the detected neutron multiplicity $M\geqslant 3$ and $M\geqslant 5$ correspondingly. As it is seen in these plots, the probability to find an overtaking signal from neutron detector in the negative range of delay times $T<0$ is about $P_{min}\approx (2-4)\cdot 10^{-4}$~$\mu$s$^{-1}\cdot$event$^{-1}$. Evidently, such pulses must be causally independent of any succeeding EAS, and it is this $P_{min}$ level which can be accepted as minimum background of random coincidences between the extensive air showers and underground neutron events.

Nevertheless, an evident and statistically reliable excess of the detected events number above $P_{min}$ is seen in both plots of figure~\ref{figiundgeasdelays} in the range of delay times between $T=0$ and $T\approx 500-800$~$\mu$s. This means that some noticeable part of detected delays can not be explained completely by accidental coincidences but must have some physical reason in EAS interaction properties. Since the primary source of underground neutron events is interaction of penetrative muons, same conclusion relates as well to the flux of the muonic component in extensive air showers.

\section{Conclusion}

Presently, the investigation results of neutron generation in the underground monitor at Tien~Shan station can be summarized as the following.
\begin{itemize}
\item[-]
The comparison of the experimental and Geant4 simulated neutron multiplicity spectra has shown that the neutron events observed underground can mostly be explained by interaction of the cosmic ray muons with internal material of neutron detector.

\item[-]
In synchronous operation of the underground neutron monitor with the Tien~Shan shower installation it was found that the origination frequency of neutron events starts to grow significantly around the knee of primary cosmic ray spectrum. Consequently, same conclusion can be made on either the average energy of EAS connected muons which are original source of neutron generation, or on the mean muon multiplicity in the above-the-knee EASs, or on both these characteristics.

\item[-]
Some peculiar EAS events were detected in which the neutron signal underground reveals itself only a few hundreds of microseconds after the trigger which is generally generated at the passage moment of an EAS front. This circumstance means an existence of corresponding delay of the EAS connected muon flux in relation to the main bulk of shower particles in these events.

\end{itemize}

\section*{Acknowledgments}
This work is supported by the grants \#BR05236201 and \#BR05236494 of IRN Program ``Fundamental and applied studies in related fields of the physics of terrestrial, near-Earth and atmospheric processes and their practical application'', and by the grant \#0118RK00800 of the Cosmic Program of Kazakhstan Republic.

\section*{References}

\begin{thebibliography}{1}
\expandafter\ifx\csname url\endcsname\relax
  \def\url#1{{\tt #1}}\fi
\expandafter\ifx\csname urlprefix\endcsname\relax\def\urlprefix{URL }\fi
\providecommand{\eprint}[2][]{\url{#2}}

\bibitem{undgour1}
Chubenko A~P, Shepetov A~L, Vildanova L~I, et al 2007
  {\em {Bull. Lebedev Phys. Inst.}\/} {\bf 34} 107--113

\bibitem{undgour2}
Chubenko A~P, Shepetov A~L, Oscomov V~V and Vildanova L~I 2008 The underground
  neutron events at {Tien-Shan} {\em {Proceedings of the 30th ICRC}\/} vol 4
  (HE part 1) (M\'{e}xico City, M\'{e}xico) pp 3--6

\bibitem{undgour2008icrcmexico}
Chubenko A~P, Shepetov A~L, Chubenko P~A,  et al 2008 The
  underground neutron calorimeter for registration of the neutron-bearing
  cosmic ray component at {Tien~Shan} {\em {Proceedings of the 30th ICRC}\/}
  vol 4 (HE1) (M\'{e}xico City, M\'{e}xico) pp 97--100

\bibitem{ontien-nim2016}
Chubenko A~P, Shepetov A~L, Antonova V~P, et al 2016 {\em {Nucl. Instrum. Methods A}\/} {\bf
  832} 158--178

\bibitem{carmichel_supermonitor}
Carmichael H and Hatton C~J 1964 {\em {Can. J. Phys.}\/} {\bf 42} 2443

\bibitem{nmn2003}
Chubenko A~P, Shepetov A~L, Antonova V~P, et al 2003 Multiplicity spectrum of {NM64} neutron supermonitor and
  hadron energy spectrum at mountain level {\em {Proceedings of the 28th
  ICRC}\/} ({Tsukuba, Japan}) pp 789--792

\bibitem{yanke2011}
Abunin A~A, Pletnikov E~V, Shchepetov A~L and Yanke V~G 2011 {\em {Bull. Russ.
  Acad. Sci. Phys.}\/} {\bf 75} 866--868

\bibitem{ontienmuons_erlykin1973}
Erlykin A~D, Kulichenko A~K, Machavariani S~K and Nikolsky S~I 1973
  Investigation of cascades, produced by high energy muons {\em Proceedings of
  the 13th {ICRC}\/} vol~3 ({Denver, Colorado, USA}: {NASA Conference
  Publication No. 2376}) p 1803 (\textit{Preprint}
  \eprint{http://adsbit.harvard.edu//full/1973ICRC....3.1803E/0001803.000.html})

\bibitem{ontien_icrc1987__e0_through_ne_ru}
Adamov D~S, Afanasjev B~N, Arabkin V~V, et al 1987 Phenomenological characteristics of {EAS}
  with {$N_e=2\cdot 10^5-2\cdot 10^7$} obtained by the modern {Tien-Shan}
  installation {``Hadron''} {\em Proceedings of the 20th {ICRC}\/} vol~5
  ({Moscow, USSR}) pp 460--463

\end{thebibliography}

\providecommand{\newblock}{}

\end{document}